\documentclass[3p,review,12pt]{elsarticle}

\usepackage{hyperref,url}
\usepackage{lineno}
\usepackage{booktabs}
\usepackage{tcolorbox}
\usepackage{amsmath}
\usepackage{amsfonts}
\usepackage{stfloats}
\usepackage{adjustbox}
\usepackage{algorithmic}
\usepackage{algorithm}
\usepackage{booktabs}
\usepackage{multirow}
\usepackage{mathtools}
\usepackage{multirow}
\usepackage{verbatim}
\usepackage{amsmath,amssymb}
\usepackage{multirow,multicol}
\usepackage{tikz}
\usepackage{lineno,hyperref}
\usepackage{subcaption,graphicx}
\usepackage{hyperref}
\usepackage{devanagari}
\usepackage{xcolor}
\usepackage{algorithm,algorithmic}
\usepackage{lscape}
\usepackage{tfrupee}
\modulolinenumbers[5]
\begin{document}
\begin{frontmatter}

\title{Investigating Target Set Reduction for End-to-End Speech Recognition of Hindi-English Code-Switching Data}

\author{Kunal Dhawan}
\ead{k.dhawan@iitg.ernet.in}
\author{Ganji Sreeram}
\ead{s.ganji@iitg.ernet.in}
\author{Kumar Priyadarshi}
\ead{k.priyadarshi}
\author{Rohit Sinha}
\ead{rsinha@iitg.ernet.in}
\address{Department of Electronics and Electrical Engineering, \\Indian Institute of Technology Guwahati, Guwahati-781039, India.}

\begin{abstract}
End-to-end (E2E) systems are fast replacing the conventional systems in the domain of automatic speech recognition. As the target labels are learned directly from speech data, the E2E systems need a bigger corpus for effective training. In the context of code-switching task, the E2E systems face two challenges: (i) the expansion of the target set due to multiple languages involved, and (ii) the lack of availability of sufficiently large domain-specific corpus. Towards addressing those challenges, we propose an approach for reducing the number of target labels for reliable training of the E2E systems on limited data. The efficacy of the proposed approach has been demonstrated on two prominent architectures, namely CTC-based and attention-based E2E networks.  The experimental validations are performed on a recently created Hindi-English code-switching corpus. For contrast purpose, the results for the full target set based  E2E system and a hybrid DNN-HMM system are also reported.\\

\end{abstract}

\end{frontmatter}

\section{Introduction}
\label{sec:intro}

\begin{figure}[]
     \centering
     \captionof{table}{Sample code-switching sentences in HingCoS corpus and their corresponding English translations.}\label{tab:ex}
     \centerline{\includegraphics[width=10cm]{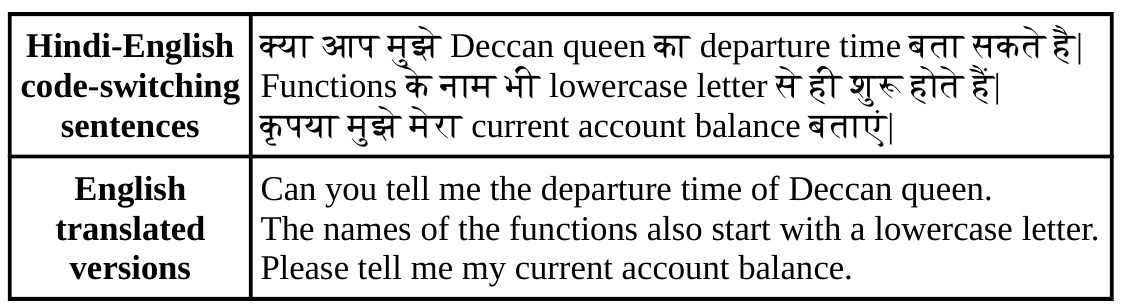}}  
\end{figure}

Code-switching is a common phenomenon in which people switch between languages for the ease of expression~\cite{Gumperz_1982_Discourse}. It has been observed that people use words of a foreign language while conversing in their native tongue so as to effectively communicate with other people~\cite{eastman1992, Myers_1992_Comparing}. The recent spread of urbanization and globalization have positively impacted the growth of bilingual/multilingual communities and hence made this phenomenon more prominent. The growth in such communities has made automatic recognition of code-switching speech an important area of interest~\cite{Lyu_2006_Speech, Bhuvanagirir_2012_Mixed, ahmed2012automatic}. In India, Hindi is the native language of around $50\%$ of its $1.32$ billion population~\cite{cen_1991}. A large portion of the remaining half, especially those residing in metropolitan cities understand the Hindi language well enough. Due to prominence in administration, law and corporate world, English language is also used by around $125$ million people in India. Thus, Indians naturally tend to use some English words within their Hindi discourse, which is referred to as Hindi-English code-switching~\cite{malhotra1980hindi, bali2014}. Despite the increasing code-switching phenomenon, the research activity in this area is somewhat limited due to lack of resources, specially for building robust code-switching ASR systems. We recently created a corpus, named as the HingCoS corpus, for addressing the data scarcity in Hindi-English code-switching domain. The initial version of the work describing the data collection for HingCoS corpus is available at~\cite{hingcos_2018}. The corpus primarily contains intra-sentential code-switching sentences and a few example ones along with their English translations are shown in Table~\ref{tab:ex}. 

End-to-end (E2E) systems are fast becoming the norm for automatic speech recognition (ASR) task. Unlike the conventional systems, the E2E systems directly model the output labels given the acoustic features. This is usually achieved by employing two techniques: (i) connectionist temporal classification (CTC) ~\cite{graves2006connectionist, graves2012sequence}, and (ii) sequence to sequence modelling with attention mechanism~\cite{graves2014towards, chorowski2014end, bahdanau2014neural, prabhavalkar2017comparison}. CTC allows us to train E2E models without the requirement of alignment between input features and output labels as required in conventional systems. It is used as cost function along with deep bi-directional long short term memory (DBLSTM) architecture to build ASR systems. Attention-based systems consist of three modules: (i) a pyramidal BLSTM network which acts as the acoustic model encoder, (ii) an attention layer which helps choose input frames to look at while making current label decision, and (iii) an LSTM network which acts as the decoder.  

\begin{figure*}[]
  \centering
    \captionof{table}{The top two rows show the default Hindi and English character sets, respectively. The proposed reduced target labels covering both Hindi and English sets are shown in the bottom row.}
  \includegraphics[width=\linewidth]{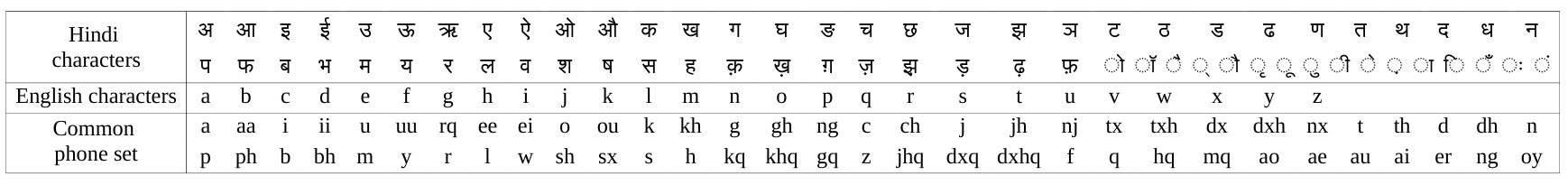} \vspace{-4mm}
  \label{fig:labels}
\end{figure*}

Conventionally, the E2E systems have been trained for characters as output labels which simplifies the process of data preparation. In~\cite{google_grapheme_phoneme}, it is shown that grapheme-based E2E ASR systems slightly outperform phoneme-based systems when a large amount of data ($>$12,500 Hrs) is used for training. Presently, building grapheme-based systems for code-switching tasks seems infeasible for two reasons. Firstly, a limited amount of data is available for code-switching tasks as yet. Secondly, the target set (output labels) in a code-switching task gets expanded in proportion to the number of languages involved. Towards addressing those constraints, we explore a target set reduction scheme by exploiting the acoustic similarity in the underlying languages of the code-switching task. This scheme is primarily intended to enhance the performances of code-switching E2E ASR systems. The validation of the proposed idea has been done on Hindi-English code-switching task using both E2E network and hybrid deep neural network based hidden Markov model (DNN-HMM).

The remainder of this paper is organized as follows: Discussion of the proposed target set reduction scheme along with a review of CTC- and attention-based E2E ASR networks is done in Section~\ref{sec:e2epar}. The experimental setup including system description is presented in Section~\ref{sec:exp}. The results are presented and discussed in Section~\ref{sec:results}. Finally, the paper is concluded in Section~\ref{sec:conclusion}.

\section{E2E Paradigms for Code-Switch ASR}
\label{sec:e2epar}
The conventional E2E ASR systems are trained directly from speech data (filterbank energies) with characters as the target labels. In the context of code-switching, a conventional E2E ASR system models the unified character set of the underlying languages. With unified character set modelling, such a system would face the following challenges:
\begin{itemize}
\item More than double expansion in the target set.
\item Enhanced confusion among the target labels.
\item Requirement of more data for reliable modelling.
\item Weakening of attention mechanism, if employed.
\end{itemize}
Towards addressing the above challenges, we first propose a novel scheme for reducing the output target labels. It is followed by the descriptions of two popular E2E architectures employed to evaluate the efficacy of the proposed scheme.
\subsection{Proposed Scheme for Reduction of Target Set}
\label{sec:proposed}
Despite the expansion of the target set in the case of code-switching E2E ASR, the phone sets corresponding to the underlying languages may have significant acoustic similarity. This fact is well known and has been exploited in the creation of a common phone set across languages~\cite{ramani2013common}. Motivated by that, we propose a scheme for target set reduction in code-switching E2E ASR task by creating common target labels based on acoustic similarity. In the following, the proposed scheme has been explained in detail in the context of Hindi-English code-switching ASR task which has been used in this work for experimentation. In principle, it can be extended to any other code-switching context as well. 

\begin{figure}[t]
  \centering
    \captionof{table}{Sample examples for the proposed common phone level labelling and existing character level labelling schemes for E2E ASR system training. Note that for the given words, the unique targets when tokenized at character level turns out to be $22$  and $12$ unique targets when tokenized using the proposed scheme.}
  \includegraphics[scale=0.6]{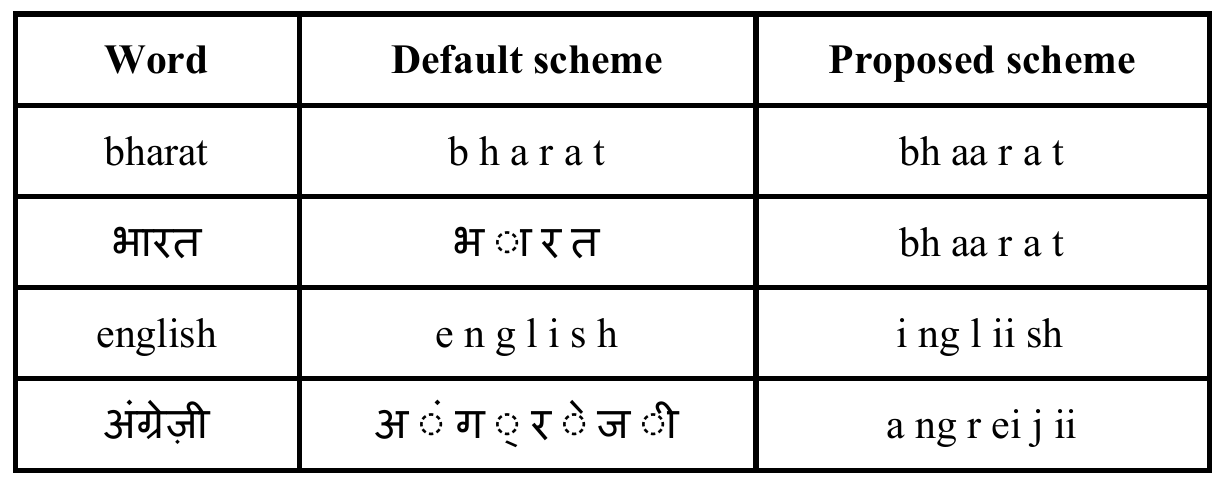}\vspace{-2mm}
  \label{fig:exs}
\end{figure}

Hindi and English languages comprise of $68$ and $26$ characters, respectively. For reference purpose, those are shown in the top two rows of Table~\ref{fig:labels}. In~\cite{ramani2013common}, a composite phone set covering major Indian languages is proposed in the context of computer processing. On a similar line, a phone set for English has been defined. Combining the labels for Hindi and English, a common phone set comprising $62$ elements is derived and is shown in the bottom row of Table~\ref{fig:labels}. Using this common phone set, a dictionary keeping the default pronunciations for all Hindi and English words in the HingCoS corpus is created. Now, the targets for acoustic modelling are derived by taking the pronunciation breakup of all Hindi and English words. A few example words along with their default character-level and the proposed common phone-level tokenizations are shown in Figure~\ref{fig:exs}. It can be observed that the considered Hindi/English words lead to $22$ unique targets when tokenized at the character level and $12$ unique targets when tokenized using the proposed scheme. 
For the Hindi-English code-switching task, the proposed approach results in $34$\% reduction in the size of the target set. The importance of this reduction gets enhanced by the fact that the availability of code-switching data is yet limited.

\subsection{CTC-based architecture}
\label{sec:CTC_DBLSTM}
CTC based E2E ASR systems consist of a DBLSTM encoder which is trained to minimize the CTC cost function. These components are described below.
\subsubsection{DBLTSM network}
Deep bidirectional long short term memory (DBLSTM) is a prominent sequence modelling architecture. It combines the advantage of multiple levels of representation that come from the use of a deep network along with long range context enabled by the use of recurrent neural networks (RNN). Conventional RNNs process sequence data from left to right, thus making use of only the previous context. In speech recognition tasks, making use of future context can be useful. Bidirectional RNNs process input data in both directions with separate hidden layers which are fed forward to the same output layer. The following equations illustrate the calculation of forward and backward activations:
\begin{align*}
\overrightarrow{h}_t = \mathcal{H}(W_{x,\overrightarrow{h}}x_t + W_{\overrightarrow{h},\overrightarrow{h}}\overrightarrow{h}_{t-1} + b_{\overrightarrow{h}})\\
\overleftarrow{h}_t = \mathcal{H}(W_{x,\overleftarrow{h}}x_t + W_{\overleftarrow{h},\overleftarrow{h}}\overleftarrow{h}_{t-1} + b_{\overleftarrow{h}})
\end{align*}
where $\overrightarrow{h}_t$ and $\overleftarrow{h}_t$ represent the forward and backward activations respectively. The other terms have their conventional meanings as defined in~\cite{graves2013hybrid}.
\noindent The output layer is given by
\begin{align*}
    y_t = W_{\overrightarrow{h},y}\overrightarrow{h}_t + W_{\overleftarrow{h},y}\overleftarrow{h}_{t} + b_y
\end{align*}
The network is trained to minimize the CTC loss function as explained in the following section.

\subsubsection{CTC cost function}
\label{sec:ctc}
CTC allows training of RNNs without requiring a prior alignment between input and output sequences. In CTC, the output softmax layer of RNN has one unit each for the targets in addition to a blank symbol $\phi$ denoting a null emission. For a given training speech example, there are as many possible alignments as there are ways of separating the labels with blanks. At every time-step, the network decides whether to emit a symbol or not.  As a result, a distribution over all possible alignments between the input and target sequences is obtained.

Finally, CTC employs a dynamic programming based forward-backward algorithm to obtain the sum over all possible
alignments and produces the probability of output sequence given a speech input. Given a target transcription $\vec{y}$ and input $\vec{x}$, the network is trained to minimize the CTC cost function:
\[\text{CTC}(\vec{x}) = - \text{log P}(\vec{y}|\vec{x})\]

\noindent Here the total probability of an output transcription
$\vec{y}$ is the sum of the probabilities of the alignments that correspond to it. So,
\[\text{P}(\vec{y}|\vec{x}) = \sum_{\vec{a}\in \vec B^{-1}(\vec{y})} \text{P}(\vec{a}|\vec{x})\]
where $\vec{a}$ corresponds to all the CTC alignments which map to required output sequence $\vec{y}$ , as represented by $\vec{y}= \vec{B}(\vec{a})$.

\begin{figure}[t]
  \centering  
  \includegraphics[scale=0.54]{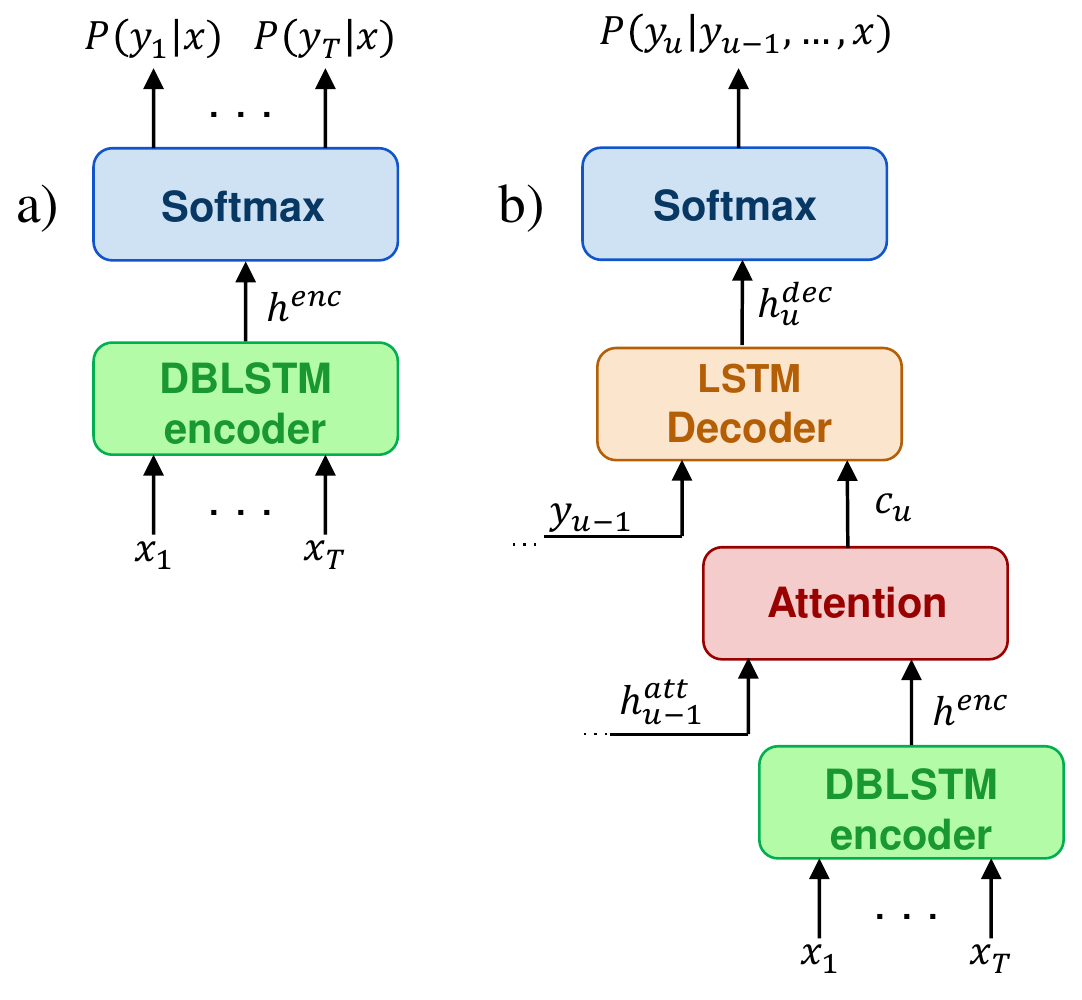}
  \caption{Block diagram of E2E Networks using: a) CTC mechanism, and b) attention mechanism.}\vspace{-2mm}
  \label{fig:ctc}
\end{figure}
\subsection{Attention-based Architecture}
\label{sec:LAS}
This model employs an encoder RNN which plays a role similar to that of an acoustic model in conventional systems, an attention layer which helps  choose the input frames to look at while making current label decision and a decoder RNN for end-to-end training of the ASR system. This architecture predicts output labels without making any independence assumption between the labels, unlike the assumption in CTC. The role of the attention layer is to select the portion of the input to be considered while making current label decision at the decoder.

In particular, we have used the listen, attend and spell~(LAS) model~\cite{las} for training the ASR system presented in this work. The LAS architecture is composed of three sub-modules: listener, attender and speller. The listener
is the acoustic encoder that transforms the original input signal $\vec{\mathbf{x}}$ into a higher level representation $\vec{\mathbf{h}}$. The AttendAndSpell function takes $\vec{\mathbf{h}}$
as input and produces a probability distribution over character/phoneme sequences whilst utilizing the attention mechanism:
\[\vec{\mathbf{h}} = \text{Listen}(\vec{\mathbf{x}})\]
\[P(y_{i}|\vec{\mathbf{x}}, y_{<i}) = \text{AttendAndSpell}(\vec{\mathbf{h}})\]

The listener/encoder is a Bidirectional LSTM network having a pyramidal structure. This reduces the number of time steps over which the attention layer has to extract relevant information and hence improves the efficacy of the attention mechanism. The speller/decoder uses an attention-based LSTM transducer and decoding is performed using a left-to-right beam search. The network is trained to optimize the following log probability:
\[\underset{\theta }{\operatorname{max}}  \sum_i \text{log} P(y_i|\mathbf{x},y*_\text{$<i$} ;\theta)\]

where $\mathbf{y}*_\text{$<i$}$ is the ground truth of the previous characters and $\theta$ represents the model parameters.

%
%

\section{Experimental Setup}
\label{sec:exp}
\subsection{Database}
In this work, the experiments are performed using the Hindi-English code switching speech corpus. This database is referred to as the {\emph {HingCoS Corpus}}\footnote{\emph{www.iitg.ac.in/eee/emstlab/HingCoS\_Database/HingCoS.html}}. An initial description of this database is available at~\cite{hingcos_2018}. It consists of 101 speakers, each of whom has been asked to sound $100$ unique code-switching sentences given to him/her. The length of those sentences varies from $3$ to $60$ words. All speech data is recorded at 8 kHz sampling rate and 16-bits/sample resolution. The database contains $9251$ Hindi-English code-switching utterances which correspond to about $25$ hours of speech data. For ASR system modelling, the database is partitioned into train, development and test sets containing $7015$, $1152$ and $2136$ sentences, respectively. To study the effect of utterance-length in decoding, three partitions of test set are created on the basis of length of utterances. Those partitions correspond to utterance-length ranges as $3$-$15$, $16$-$25$, and $26$-$60$ words and are referred to as Test1, Test2, and Test3, respectively. So obtained, Test1, Test2, and Test3 data sets consist of $957$, $719$, and $460$ utterances, respectively. The unified character set modelling case comprises of $95$ targets ($26$ English characters,  $68$ Hindi characters, and a word separator). In contrast, the proposed scheme reduces that to $63$ targets ($62$ common phones and a word separator). In this work, we contrast the performances of the proposed reduced target set based E2E ASR systems with those of unified character set based ones.


\subsection{System Description}
The E2E models developed in this work are trained using the Nabu toolkit~\cite{nabu_2017}, which 
is based on TensorFlow$^\circledR$. For contrast purpose, the DNN-HMM systems have also been trained and evaluated using Kaldi toolkit~\cite{povey2011kaldi}. The parameter setting used for analyzing the speech data include window length of $25$ ms, window shift of $10$ ms, and pre-emphasis factor of $0.97$. The $26$-dimensional features comprising log filter-bank energies are used for developing E2E systems. It is to be noted that, the E2E systems are optimized for the reduced target set and the same parameters have been used for the unified character set systems. The remaining details of the above mentioned systems are presented next.


\begin{table*}[]
\centering
\caption {Evaluation of attention and CTC based E2E systems developed using both reduced and unified target sets on Hindi-English code-switching data. The performances of reduced and unified target set based systems are measured using phone error rate (PER) and character error rate (CER), respectively. The performances for DNN-HMM system on those tasks are also given for reference purpose.}\label{tab:results}
\begin{tabular}{|c|c|c|c|c|c|c|c|c|}
\hline
\multirow{2}{*}{\textbf{Model}} & \multicolumn{2}{c|}{\textbf{Test1}} & \multicolumn{2}{c|}{\textbf{Test2}} & \multicolumn{2}{c|}{\textbf{Test3}} & \multicolumn{2}{c|}{\textbf{Average}} \\ \cline{2-9} 
                                & \textbf{PER}      & \textbf{CER}     & \textbf{PER}      & \textbf{CER}     & \textbf{PER}      & \textbf{CER}     & \textbf{PER}      & \textbf{CER}      \\ \hline \hline
\textbf{Attention-based E2E}    & 21.01             & 33.69            & 21.06             & 34.80            & 23.70             & 39.38            & 21.92             & 35.96             \\ \hline
\textbf{CTC-based E2E}          & 32.91             & 35.82            & 28.89             & 32.85            & 28.33             & 33.87            & 30.04             & 34.18             \\ \hline \hline
\textbf{DNN-HMM}                & 48.21             & 48.74            & 47.85             & 48.17            & 47.88             & 48.62            & 47.98             & 48.51             \\ \hline
\end{tabular}
\end{table*}

\begin{figure*}[]
\captionof{table}{Sample decoded outputs for E2E code-switching ASR systems developed using reduced and unified target sets. The errors have been highlighted in bold. Note that, the symbol \textbf{`\_'} is used to mark separation between the words.}\label{tab:examples}
\centering
{\includegraphics[scale=0.65]{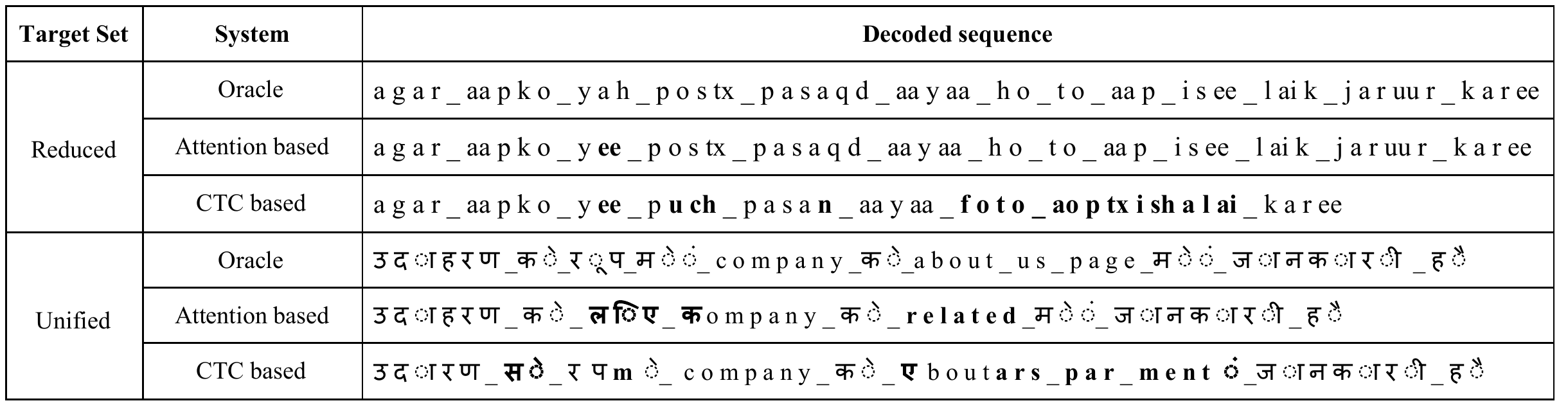}}\vspace{-3.1mm}
\end{figure*}

\subsubsection{Attention-based E2E model}
\label{sec:att_desc}
The architectural details of the LAS model are as follows.  The encoder has $3$ pyramidal DBLSTM layers with $512$ units in each layer. The pyramidal step size is kept as $2$ and the dropout rate in training is set to $0.5$. The LSTM decoder consists of $2$ layers with $512$ units in each layer. The dropout rate for the LSTM decoder is also set to $0.5$. Loss function used for training is the average cross-entropy loss and Gaussian noise with $\sigma = 0.6$ is added to the data while training. We have employed the beam-search decoder with beam width set as $16$. The model is trained for $400$ epochs with a batch size of $32$ and learning rate decay set to $0.1$.



\subsubsection{CTC-based E2E model}
This modelling paradigm involves a DBLSTM network as the encoder which consists of 4 layers and 256 units in each layer with dropout rate set to $0.5$. The decoder utilizes CTC loss function as discussed in Section~\ref{sec:ctc}. Gaussian noise with $\sigma = 0.6$ is added to the speech data for modelling robustness. In model training, the number of epochs is set as $250$ and the mini-batch size is set to $32$. 


\subsubsection{DNN-HMM model}
The DNN-HMM acoustic model contains $5$ hidden layers and $1024$ nodes in each layer. The hidden nodes use \emph{tanh} as the non-linearity. First, $13$-dimensional MFCC features are spliced across $\pm3$ frames to produce $91$-dimensional feature vectors, which are then projected to $40$ dimensions by applying linear discriminant analysis. These $40$-dimensional feature vectors are used for training the DNN-HMM acoustic model. The model is run for $20$ epochs with a batch size of $128$.

\section{Results and Discussion}
\label{sec:results}

For the unified target set case, the performances are measured in terms of the character error rate (CER). Whereas, for the reduced target set case, we have used the phone error rate (PER) as the measure. For proper evaluation, both attention and CTC based E2E ASR systems are developed using reduced and unified target sets and their performances are reported in Table~\ref{tab:results}. It can be observed that with proposed reduction in target set, all explored E2E systems yield significantly improved recognition performance (i.e., target error rate) over the corresponding unified target set based systems. Interestingly, this trend is carried over all the three test sets as defined earlier. On comparing the reduced target set systems, we note that the attention-based E2E ASR system has outperformed. Whereas, the CTC-based E2E system has yielded slightly better CER for the unified target set modelling case.

It is worth emphasizing that with more reduction in the target set further improvement in PERs could be achieved in reference to CERs. But any such reduction would be counterproductive if we can not derive accurate word sequences given the output hypotheses in terms of those reduced target labels. That criterion is very much satisfied by the proposed phone based reduction of the target set in the case of code-switching speech. On the other hand, for unified target set based E2E ASR systems, the decoded outputs may comprise of cross-language character insertions due to acoustic similarity. Towards illustrating that, we show  a few example decoded sequences for both reduced and unified target set based E2E systems in Table~\ref{tab:examples}. From that table, we can note that the decoded sequence for the attention-based E2E system exhibits better sequence modelling as well as word boundary marking in comparison to that of CTC-based system.  This trend is attributed to the ability of attention-based E2E network to utilize all the previous decoded labels along with the current input while making decisions.


\section{Conclusions}
In this work, we present a novel target label reduction scheme for training the E2E code-switching ASR systems. The systems employing the reduced targets are shown to outperform the unified target based systems. It has been demonstrated that the attention based E2E system trained with reduced target set achieves the best averaged target (phone) error rate. In the future, we aim to incorporate language information in E2E code-switching ASR systems developed in this work under the paradigm of multi-task learning. 

\label{sec:conclusion}

\bibliography{ASR}

\end{document}